\documentclass[twocolumn]{aastex631}
\usepackage{graphicx}
\usepackage{amsmath}
\usepackage{color}

\newcommand{\DEL}[1]{}
\newcommand{\Msol}{\hbox{M$_\sun$}}

\newcommand{\zsp}{\hbox{$z_{\rm spec}$}}
\newcommand{\zph}{\hbox{$z_{\rm phot}$}}

\begin{document}

\title{The Distribution of Quenched Galaxies in the Massive $\boldsymbol{z=0.87}$ Galaxy Cluster El Gordo}

\author[0000-0002-9984-4937]{Rachel Honor}
\affiliation{School of Earth and Space Exploration, Arizona State University, Tempe, AZ 85287-1404, USA}

\author[0000-0003-3329-1337]{Seth H. Cohen}
\affiliation{School of Earth and Space Exploration, Arizona State University, Tempe, AZ 85287-1404, USA}

\author[0000-0001-6650-2853]{Timothy Carleton}
\affiliation{School of Earth and Space Exploration, Arizona State University, Tempe, AZ 85287-1404, USA}

\author[0000-0002-9895-5758]{S. P. Willner} 
\affiliation{Center for Astrophysics \textbar\ Harvard \& Smithsonian, 
60 Garden Street, Cambridge, MA, 02138, USA}

\author[0000-0001-7411-5386]{Maria del Carmen Polletta} 
\affiliation{INAF – Istituto di Astrofisica Spaziale e Fisica Cosmica Milano, Via A. Corti 12, I-20133 Milano, Italy}

\author[0000-0001-8156-6281]{Rogier A. Windhorst}
\affiliation{School of Earth and Space Exploration, Arizona State University,
Tempe, AZ 85287-1404, USA}

\author[0000-0001-7410-7669]{Dan Coe} 
\affiliation{Space Telescope Science Institute, 3700 San Martin Drive, Baltimore, MD 21218, USA}
\affiliation{Association of Universities for Research in Astronomy (AURA) for the European Space Agency (ESA), STScI, Baltimore, MD 21218, USA}
\affiliation{Center for Astrophysical Sciences, Department of Physics and Astronomy, The Johns Hopkins University, 3400 N Charles St. Baltimore, MD 21218, USA}

\author[0000-0003-1949-7638]{Christopher J. Conselice} 
\affiliation{Jodrell Bank Centre for Astrophysics, Alan Turing Building,
University of Manchester, Oxford Road, Manchester M13 9PL, UK}

\author[0000-0001-9065-3926]{Jose M. Diego}
\affiliation{Instituto de Física de Cantabria (CSIC-UC), Avda. Los Castros s/n, 39005 Santander, Spain}

\author[0000-0001-9491-7327]{Simon P. Driver} 
\affiliation{International Centre for Radio Astronomy Research (ICRAR) and the
International Space Centre (ISC), The University of Western Australia, M468,
35 Stirling Highway, Crawley, WA 6009, Australia}

\author[0000-0002-9816-1931]{Jordan C. J. D'Silva} 
\affiliation{International Centre for Radio Astronomy Research (ICRAR) and the
International Space Centre (ISC), The University of Western Australia, M468,
35 Stirling Highway, Crawley, WA 6009, Australia}
\affiliation{ARC Centre of Excellence for All Sky Astrophysics in 3 Dimensions
(ASTRO 3D), Australia}

\author[0000-0002-7460-8460]{Nicholas Foo}
\affiliation{School of Earth and Space Exploration, Arizona State University, Tempe, AZ 85287-1404, USA}

\author[0000-0003-1625-8009]{Brenda L. Frye}
\affiliation{Department of Astronomy/Steward Observatory, University of Arizona, 933 N. Cherry Avenue, Tucson, AZ 85721, USA}

\author[0000-0001-9440-8872]{Norman A. Grogin} 
\affiliation{Space Telescope Science Institute,
3700 San Martin Drive, Baltimore, MD 21218, USA}

\author[0000-0001-6145-5090]{Nimish P. Hathi}
\affiliation{Space Telescope Science Institute,
3700 San Martin Drive, Baltimore, MD 21218, USA}

\author[0000-0003-1268-5230]{Rolf A. Jansen} 
\affiliation{School of Earth and Space Exploration, Arizona State University,
Tempe, AZ 85287-1404, USA}

\author[0000-0001-9394-6732]{Patrick S. Kamieneski}
\affiliation{School of Earth and Space Exploration, Arizona State University, Tempe, AZ 85287-1404, USA}

\author[0000-0002-6610-2048]{Anton M. Koekemoer} 
\affiliation{Space Telescope Science Institute,
3700 San Martin Drive, Baltimore, MD 21218, USA}

\author[0009-0001-7446-2350]{Reagen Leimbach}
\affiliation{Department of Astronomy/Steward Observatory, University of Arizona, 933 N. Cherry Avenue, Tucson, AZ 85721, USA}

\author[0000-0001-6434-7845]{Madeline A. Marshall} 
\affiliation{National Research Council of Canada, Herzberg Astronomy \&
Astrophysics Research Centre, 5071 West Saanich Road, Victoria, BC V9E 2E7,
Canada}
\affiliation{ARC Centre of Excellence for All Sky Astrophysics in 3 Dimensions
(ASTRO 3D), Australia}

\author[0000-0002-6150-833X]{Rafael {Ortiz~III}} 
\affiliation{School of Earth and Space Exploration, Arizona State University,
Tempe, AZ 85287-1404, USA}

\author[0000-0003-3382-5941]{Nor Pirzkal} 
\affiliation{Space Telescope Science Institute,
3700 San Martin Drive, Baltimore, MD 21218, USA}

\author[0000-0003-4223-7324]{Massimo Ricotti}
\affiliation{Department of Astronomy, University of Maryland, College Park, 20742, USA}

\author[0000-0003-0429-3579]{Aaron S. G. Robotham} 
\affiliation{International Centre for Radio Astronomy Research (ICRAR) and the
International Space Centre (ISC), The University of Western Australia, M468,
35 Stirling Highway, Crawley, WA 6009, Australia}

\author[0000-0001-7016-5220]{Michael J. Rutkowski}
\affiliation{Department of Physics and Astronomy, Minnesota State University, Mankato, MN 56001, USA}

\author[0000-0003-0894-1588]{Russell E. Ryan, Jr.} 
\affiliation{Space Telescope Science Institute,
3700 San Martin Drive, Baltimore, MD 21218, USA}

\author[0000-0002-5319-6620]{Payaswini Saikia}
\affiliation{Center for Astrophysics and Space Science (CASS), New York University Abu Dhabi, PO Box 129188, Abu Dhabi, UAE}

\author[0000-0002-7265-7920]{Jake Summers} 
\affiliation{School of Earth and Space Exploration, Arizona State University,
Tempe, AZ 85287-1404, USA}

\author[0000-0001-9262-9997]{Christopher N. A. Willmer} 
\affiliation{Steward Observatory, University of Arizona,
933 N Cherry Ave, Tucson, AZ, 85721-0009, USA}

\author[0000-0001-7592-7714]{Haojing Yan} 
\affiliation{Department of Physics and Astronomy, University of Missouri,
Columbia, MO 65211, USA}



\begin{abstract}
El Gordo (ACT-CL J0102$-$4915) is a massive galaxy cluster with two major mass components at redshift $z=0.87$. Using SED fitting results from JWST/NIRCam photometry, the fraction of quenched galaxies in this cluster was measured in two bins of stellar mass: $9<\log{({M_*}/\mathrm{M}_{\odot})}<10$ and $10\leq\log{({M_*}/\mathrm{M}_{\odot})}<12$. While there is no correlation between the quenched fraction and angular separation from the cluster's overall center of mass, there is a correlation between the quenched fraction and angular separation from the center of the nearest of the two mass components for  the less-massive galaxies. This suggests that environmental quenching processes are in place at $z\sim1$, and that dwarf galaxies are more affected by those processes than massive galaxies.
\end{abstract}



\section{Introduction} \label{sec:intro}

Observations have long shown a bimodal galaxy population, with quiescent spheroidals relatively more abundant among more massive galaxies, and those in more dense environments \citep{1980ApJ...236..351D,2005ApJ...629..143B}. The mechanisms by which star-forming galaxies transition to quiescent galaxies is an active area of research in modern astrophysics (e.g., \citealt{2022Univ....8..554A,2003MNRAS.341...54K,2004MNRAS.353..713K,2008MNRAS.391..481S,2009ApJ...707..250M,2010ApJ...721..193P,2015Natur.521..192P,2016ApJ...833....1L,1977MNRAS.179..541R,1998A&A...331L...1S,2006MNRAS.368....2D,2006MNRAS.365...11C,2010MNRAS.402.1693H,2017MNRAS.464.1077W,2020ApJ...897..102C,2022ApJ...926..134T}). 

Suppression of star formation driven by internal mechanisms is referred to as internal quenching, whereas when it is due to processes linked to 
galaxy density and host halo mass it is called environmental quenching (\citealt{2010ApJ...721..193P}, \citealt{2013MNRAS.432..336W}). For massive halos, feedback provided by supernovae, AGN activity, stellar winds, and (at very early times)
reionization (see \citealt{1986ApJ...303...39D}; \citealt{2006MNRAS.368....2D}) are thought to heat the gas in massive halos, preventing new gas accretion and star formation.
For galaxies in groups and clusters, dynamical and hydrodynamic interactions between satellite galaxies and their hosts play a major role in driving quenching through processes such as ram-pressure stripping (\citealt{1972ApJ...176....1G}, \citealt{2017IAUS..321..202P}), strangulation (\citealt{1980ApJ...237..692L}, \citealt{1972ApJ...176....1G}), tidal interactions (\citealt{1998ApJ...495..139M}), and gas-rich mergers (e.g., \citealt{2010MNRAS.402.1693H}, \citealt{2012MNRAS.427.1816G}, \citealt{2010ApJ...721..193P}). While a few studies have reported elevated quenched fractions with increased proximity to neighbors among galaxies in clusters at intermediate redshifts (e.g., \citealt{2017ApJ...841L..22G,2021ApJ...914....7G,2022MNRAS.515.5479B}), the role of the environment in quenching star formation at intermediate redshifts and in massive halos requires further investigation, as there are several processes that can affect star formation, and observational constraints are largely missing (\citealt{2015Sci...348..314T}, \citealt{2019ApJ...872...50L}, \citealt{2019ApJ...870...19G}, \citealt{2020MNRAS.493L..39E}).

A commonly invoked mechanism driving the environmental quenching of galaxies is violent stripping, including ram-pressure and viscous stripping of the galaxy's gas supply as it moves through the hot intracluster medium (ICM) (\citealt{1972ApJ...176....1G,2016MNRAS.463.1916F,2022A&ARv..30....3B}). 

The difference in quenched fraction in field galaxies versus cluster galaxies has been explored in multiple works. At $z<1$, clusters have lower star-forming fractions than the field \citep{2020MNRAS.493.5987O, 1980ApJ...236..351D, 1994ApJ...430..107D}. \citet{2004MNRAS.353..713K} demonstrated that the star-formation history of galaxies is very environment-dependent with a stronger dependence for galaxies with masses below $3\times 10^{10}$ $\mathrm{M}_\odot$. \citet{2013MNRAS.432..336W} showed that the quiescent fraction for clusters decreases with increased redshift and that this trend is stronger in low-mass galaxies. Other works (e.g., \citealt{1984ApJ...285..426B, 2006ApJ...642..188P, 2011MNRAS.413..996M, 2011ApJ...739L..40R}) have found the same trend.

Observations show a correlation between the star formation rate (SFR) and stellar mass ($M_*$) of star-forming galaxies called the star-forming main sequence (\citealp[SFMS, e.g.,][]{2009AIPC.1201...45C, 2004MNRAS.351.1151B, 2007ApJ...670..156D, 2007A&A...468...33E, 2007ApJ...660L..43N, 2012ApJ...754L..29W,Popesso2023}). The evolution of the SFMS provides insight to the interplay of star formation and factors which inhibit star formation, such as feedback from supernovae and AGN, over cosmic time. \citet{2014ApJS..214...15S} showed that this correlation is redshift-dependent with the normalization increasing with increasing redshift. 

In order to better understand environmental quenching, rich cluster environments are needed as subjects. The cluster ACT-CL J0102$-$4915 \citep{2011ApJ...737...61M}, also known as El Gordo, is remarkably massive for its redshift of $z=0.87$ \citep{2012ApJ...748....7M}. Strong-lensing measurements have found that the halo mass of the cluster is ${\sim} 10^{15}\,\mathrm{M_\odot}$ \citep{2013ApJ...770L..15Z,2018ApJ...859..159C,2020ApJ...904..106D,2023A&A...678A...3C}. Recent JWST studies have confirmed El Gordo's high mass, although the exact value is still disputed \citep{2023A&A...672A...3D,2023ApJ...952...81F}. El Gordo has two main mass components: one in the northwest (NW) and one in the southeast (SE) \citep{2023ApJ...952...81F}. The mass ratio between components appears to be near 1:1, although some studies (e.g., \citealt{2012ApJ...748....7M,2014ApJ...785...20J}) favor the NW component while others (e.g., \citealt{2023ApJ...952...81F,2013ApJ...770L..15Z,2018ApJ...859..159C,2020ApJ...904..106D,2021ApJ...923..101K,2023A&A...678A...3C,2023A&A...672A...3D}) favor the SE component. In this work, we adopt the coordinates of the mass peaks reported by the \citet{2023ApJ...952...81F} lens model.

The presence of two near-equal mass components argues that El Gordo represents a merger of two galaxy clusters, although the literature has not yet converged on its dynamical state. \citet{2015ApJ...800...37M} showed that it appears to be post-first passage, and \citet{2015MNRAS.453.1531N} found that it now appears to be post-maximum separation and is in the return phase. There is not a consensus on whether the collision was head-on or off-axis \citep{2018ApJ...855...36Z}.

Notably, El Gordo has medium-deep JWST/NIRCAM imaging as part of the PEARLS GTO program. This, combined with its high redshift and extreme mass, means that it is well-suited to study environmental quenching in high-z galaxy clusters.
This work utilizes JWST/NIRCam data to study this relationship, specifically the effects of stellar mass and distance from the cluster mass peaks on quenching star formation. The high resolution of JWST/NIRCam allows us to include low-mass galaxies in our study of quenching in the El Gordo galaxy cluster.

This paper is organized as follows. Section 2 introduces the JWST NIRCam observations of El Gordo and photometry. Section 3 describes our cluster-member selection process using a color--color diagram and photometric redshifts. SED fitting is discussed in Section~4. Section~5 discusses the implications of the results, and Section~6 summarizes them.  All magnitudes are AB \citep{1983ApJ...266..713O}, and the cosmological model used in this work is a flat $\Lambda\mathrm{CDM}$ model with $H_0=67.66\,\mathrm{km}\,\mathrm{s}^{-1}\,\mathrm{Mpc}^{-1}$, $\Omega_{M,0}=0.31$, and $T_{\rm CMB,0}=2.7255\,\mathrm{K}$ \citep{2020A&A...641A...6P}.
With those parameters, the angular scale at $z=0.87$ is 7.94~kpc~arcsec$^{-1}$.

\section{Observations} \label{sec:style}

\subsection{JWST/NIRCam Observations} \label{subsec:pearls_observations}
The primary data used in this work are JWST/NIRCam images from the Prime Extragalactic Areas for Reionization and Lensing Science (PEARLS) survey (PI: Windhorst, Program ID 1176). The El Gordo field was observed 2022 July 29 with the F090W, F115W, F150W, F200W, F277W, F356W, F410M, and F444W filters.  NIRCam has two modules, each observing an $\sim$2\farcm2 square field of view (FoV) with the two FoVs separated by $\sim$44\arcsec.  For the El Gordo observations, the ``B'' module was centered on the cluster, and the ``A'' module observed a sky field to the north \citep[][their Fig.~1]{2023ApJ...952...81F}.

The NIRCam images were reduced by the PEARLS team as described by \citet{2023AJ....165...13W}. In summary, data were retrieved from the Mikulski Archive for Space Telescopes (MAST), calibrated with the JWST pipeline\footnote{\url{https://github.com/spacetelescope/jwst}} \citep{2023zndo...6984365B}, corrected for $1/f$ noise, wisps, snowballs, and detector-level offsets using \texttt{ProFound} \citep{2018MNRAS.476.3137R, 2023ascl.soft12020R} and the processing software \texttt{JumProPe} \citep{2023ApJ...959L..18D}.\footnote{\url{https://github.com/JordanDSilva/JUMPROPE}} The resulting calibrated images were drizzled onto the GAIA-DR3 reference frame\footnote{\url{https://www.cosmos.esa.int/web/gaia/dr3}} with 0\farcs030 pixels, following procedures originally described by \citet{2011ApJS..197...36K} and updated for JWST observations. Version $1.11.2$ of the pipeline was used with \texttt{pmap\char`_1100}.  Because of dithering, the final images have FoVs $\sim$2\farcm3 on a side with $\sim$37\arcsec\ separation between modules. The specific observations analyzed for this work can be accessed via doi: \url{10.17909/w7n4-qz71}.

\label{subsec:photometry}
To detect galaxies, \texttt{Source\-Extractor} \citep{1996A&AS..117..393B} was run in dual image mode using F444W as the detection image. This corresponds to emission at rest-frame 2.4\,\micron, which is a good tracer of stellar mass \citep[e.g.,][]{Bell2003}. The minimum number of pixels above the threshold (\texttt{DETECT\char`_MINAREA}) was set to 4, \texttt{DETECT\char`_THRESH} to $1.5 \sigma$, \texttt{ANALYSIS\char`_THRESH} to $1.5 \sigma$, the number of deblending sub-thresholds (\texttt{DEBLEND\char`_NTHRESH}) to 32, and the minimum contrast parameter for deblending (\texttt{DEBLEND\char`_MINCONT}) to 0.06. These parameters produced 2,593 sources in the El Gordo NIRCam field and 2,671 sources in the sky field (5,264 total). We adopted \texttt{MAG\char`_AUTO} as the observed magnitudes and \texttt{MAGERR\char`_AUTO} as the uncertainty in magnitude in all eight NIRCam filters. The catalog reaches 5$\sigma$ depths of  27.9, 28.0, 28.0, 28.3, 29.4, 29.4, 28.8, and 29.1~mag for F090W, F115W, F150W, F200W, F277W, F356W, F410M, and F444W, respectively. A single orbit of HST F435W data for El Gordo exists \citep{2019ApJ...884...85C}, but it is too shallow to allow for significantly improved SED fits.

\subsection{Spectroscopic Redshifts}
In addition to the JWST observations, we make use of spectroscopic observations of El Gordo from \citet{2012ApJ...748....7M} and \citet{2023A&A...678A...3C}. \citeauthor{2012ApJ...748....7M} used the Very Large Telescope (VLT) FORS2 instrument in 2011 January with the GRIS 300+11 grism. The VLT spectra have resolution $R\equiv \lambda/\Delta\lambda\sim660$
and span  $\sim$4000--8000~\AA. 
\citeauthor{2012ApJ...748....7M} measured redshifts with the RVSAO/XCSAO IRAF package \citep{1998PASP..110..934K} and spectral templates from the Sloan Digital Sky Survey (SDSS) Data Release 7 \citep{2009ApJS..182..543A}. \citet{2023A&A...678A...3C} obtained MUSE 
observations with three $\approx$2.3-hour pointings  between 2018 December and 2019 September (ESO Program ID 0102.A-0266, P.I.: G. B. Caminha). These observations cover 4700--9350~\AA\ with a gap at 5805--5965~\AA. 
\citeauthor{2023A&A...678A...3C} measured redshifts either from spectral templates or by finding emission lines.  Matching the two redshift lists to the F444W catalog (with match radius 0\farcs5) found spectroscopic redshifts for 206 catalog sources.

\begin{figure*}[tb!]
    \plotone{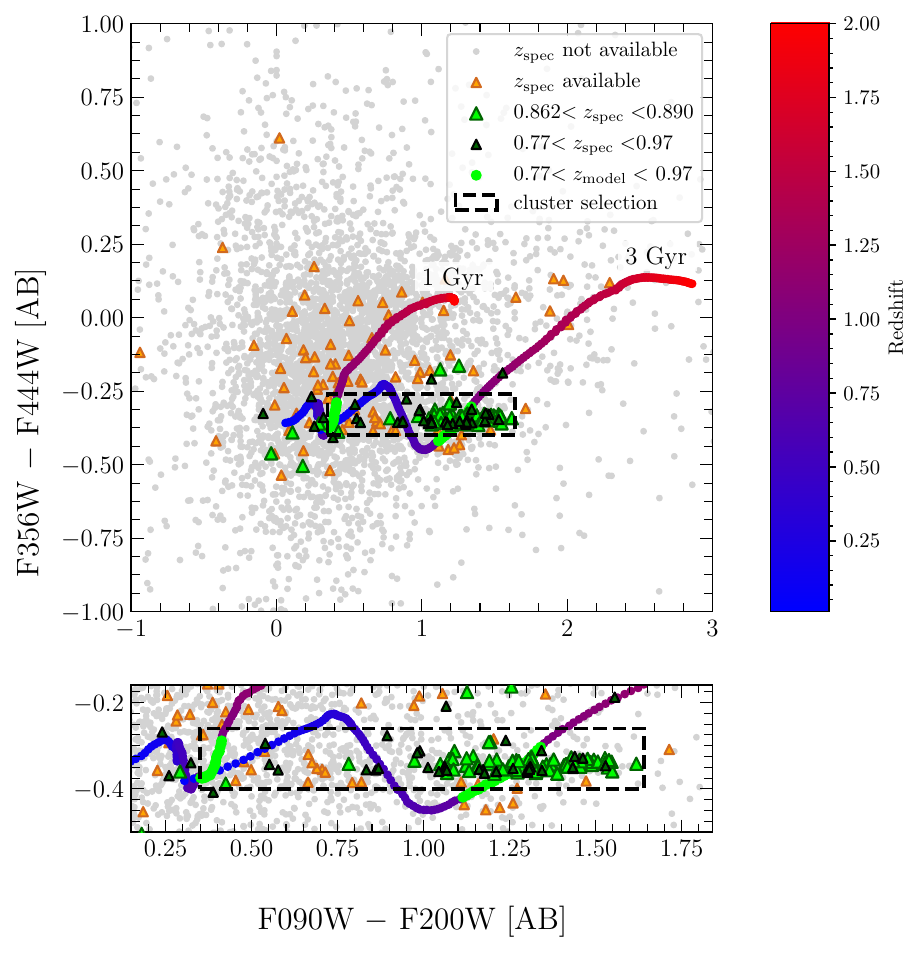}
    \caption{Color--color diagrams with (F356W$-$F444W) versus (F090W$-$F200W). The blue points represent objects without spectroscopic redshifts, and the orange triangles represent the 206 objects with spectroscopic redshifts  \citep{2012ApJ...748....7M,2023A&A...678A...3C}. Redshift tracks for two simple stellar populations \citep{2003MNRAS.344.1000B} are plotted, one with age 1~Gyr track and one with age 3~Gyr. Redshifts along each track correspond to the color bar at right, and each track is highlighted green for $0.77<z<0.97$. The lime green triangles show the 80 objects with $0.862<z_\mathrm{spec}<0.890$ \citep{2023ApJ...952...81F}. The dark green triangles show the 38 objects outside this  range but within $0.80<z_\mathrm{spec}<1.0$. The top panel shows a wide color range with the black dashed box marking the cluster-member selection region. The bottom panel shows a
    zoomed-in view of the cluster selection box.}
    \label{fig:clust_sep}
\end{figure*}

\section{Cluster Member Selection}

The spectroscopic redshifts, \zsp, identify 80 members of El Gordo, with redshifts $0.862\le z\le 0.890$, but a much larger sample is needed to draw any statistically-significant conclusions about the effects of stellar mass or distance from the nearest mass peak on the quenched fraction.
Cluster members can be identified by
the 1.6\,\micron\ bump, which comes from the H$^-$ opacity minimum.  \citet{2002AJ....124.3050S} showed that this feature determines photometric redshifts of a wide range of stellar populations.  At El Gordo's redshift, the feature is observed at 3.0\,\micron, within the range well sampled by NIRCam.

We used two complementary methods for using the 1.6\,\micron\ bump to identify cluster members.  The simpler one is a color--color diagram as shown in Figure~\ref{fig:clust_sep}. In practice, galaxies with $\rm0.35\le[F090W]-[F200W]<1.65$ and $\rm-0.4\le[F356W]-[F444W]<-0.26$ were considered cluster members by this method. There are 672 galaxies in total which satisfy this color criterion, including 74 of the 80 spectroscopic cluster members and 51 galaxies that spectroscopy shows to be non-members. We will refer to the latter category of galaxies, those which are shown by \zsp\, not to be members of El Gordo but are selected as such by another method, as ``impostors." Most of these impostors are at $0.6 \la z\la0.7$ and could be part of the $z\approx0.62$ overdensity present in the El Gordo field \citep{2023ApJ...952...81F}. The stellar-population tracks in Figure~\ref{fig:clust_sep} show that these $z\approx0.6$ interlopers are to be expected when using only this color--color diagram for cluster selection, meaning a more reliable selection method is required.

\begin{figure}
    \centering
    \includegraphics[width=0.99\linewidth]{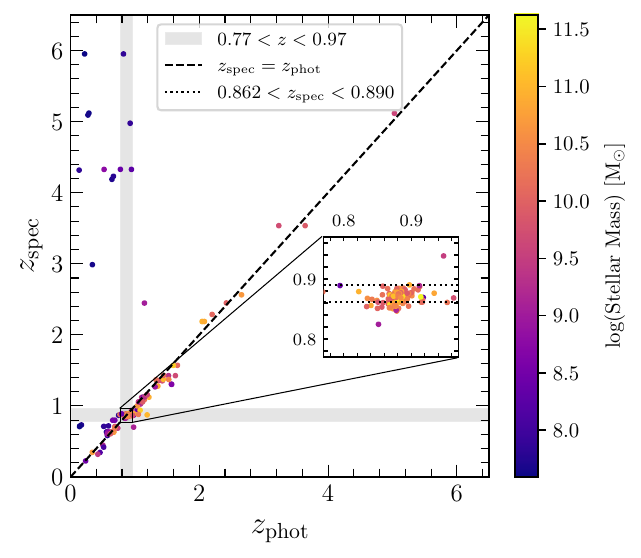}
    \caption{A comparison of the \zsp\ and \zph, colored by stellar mass. The inset box shows the area $0.77<z<0.97$ to highlight the sources within $0.1$ of the cluster redshift, $z=0.87$. The dashed line shows $\zsp = \zph$, and the dotted lines inside the smaller box show the exact cluster redshift range, $0.862<\zsp<0.890$.}
    \label{fig:zs_zp}
\end{figure}

To refine the selection of cluster members in this work, photometric redshifts, \zph, were used in conjunction with the aforementioned color--color selection. These photometric redshifts take advantage of all available photometry. For this work, \zph\ for all catalog sources came from the template-fitting program
\texttt{EAZY} \citep{2008ApJ...686.1503B}. The templates used were the default 12 Flexible Stellar Population Synthesis (FSPS) models. The \texttt{Z\char`_STEP} parameter was set to 0.01, \texttt{Z\char`_MIN} to 0.01, and \texttt{Z\char`_MAX} to 12. We did not apply a prior on \zph\ because the cluster redshift distribution is different from the field's redshift distribution. We adopted \texttt{EAZY}'s  ``maximum-likelihood redshift"  \texttt{z\char`_phot} as the \zph\ estimate and \texttt{z\char`_phot\char`_chi2} as the measure of how well the observed SED matched \texttt{EAZY}'s templates. Tests of the validity of \texttt{EAZY}'s \zph\ estimates included a $\chi ^2$ histogram to ensure the $\chi ^2$ values were reasonable, a comparison of \zph\ vs. \zsp , shown in Figure \ref{fig:zs_zp}, and visual inspection of the \texttt{EAZY}-generated probability distributions and SED fits for individual objects.

The \texttt{EAZY} selection requirement $\zph= 0.87\pm 0.10$ identified 75 of the 80 spectroscopic cluster members and 38 impostors. Of these impostors, 4 have $\zsp > 4$, outside \texttt{EAZY}'s optimum range \citep{2008ApJ...686.1503B} and thus more liable to be assigned an incorrect \zph.  While this type of impostor makes up only a small fraction of the galaxies with spectroscopy, they probably constitute a larger fraction of fainter galaxies without spectroscopy. Another 34 impostors have $0.77 \le \zsp\le 0.97$ but $\zsp<0.862$ (32 cases) or $\zsp>0.890$ (2 cases), where nothing except spectroscopy can rule them out as cluster members. For the 206 galaxies with available \zsp\ values, we default to these values when determining cluster membership. This means that the impostors are not considered cluster members in practice and cluster members which were not selected are added to the selected sample.

We adopted the most conservative sample of cluster members by requiring members to be identified at the cluster redshift using both color--color and photometric redshifts.  The main advantage of using the combination of these methods is that no objects with $\zsp >4$ or $\zsp \approx 0.6$ are selected as members.
The combined selection yielded 343 El Gordo members, 308 in the cluster field (NIRCam Module B) and 35 in the sky field (NIRCam Module A). Assuming all the objects in the sky field are non-members gives a lower bound of 90\% (308/343) on the reliability estimate. 

This selection includes 105 of the 206 galaxies with available spectroscopy. Of these 105 galaxies, 32 are impostors due to having either \zsp$<0.862$ (30 cases) or \zsp$>0.890$ (2 cases). This 70\% (73/105) reliability is lower than our estimate from the sky field. However, we note that all 32 impostors are very close to being within the cluster redshift range. All 30 of the galaxies with too low a redshift are within 0.015 of the lower bound ($z=0.862$), and the 2 galaxies with too high a redshift are 0.04 and 0.00017 from the upper bound ($z=0.890$). The large portion of galaxies with redshifts just below the cluster range could suggest an infalling group, though this is only speculative with the information available. At the very least, the impostors in this sample are likely associated with the cluster in some way, as opposed to the $\zsp > 4$ and $\zsp \approx 0.6$ impostors found with previous selection methods. Assuming they are all close enough to the cluster redshift range to be infalling cluster members gives us an upper bound of 100\% on our reliability estimate.

The rest of the selected galaxies with \zsp\ represent 73 of the 80 galaxies with $0.862<$\zsp$<0.890$, suggesting a completeness value of 91\% (73/80). However, galaxies with \zsp\ are relatively bright, so the completeness is probably worse for fainter galaxies. The impostors with \zsp\ outside the cluster redshift range were excluded from analysis and those 7 members not selected were included for a final sample of 318 galaxies in total, with 283 in the cluster field and 35 in the sky field. 

\DEL{
After estimating the photometric redshifts for all objects using \texttt{EAZY}, this object catalog was matched with catalogs from \citet{2012ApJ...748....7M} and \citet{2023A&A...678A...3C} that include spectroscopic redshift measurements. Sources within 0\farcs5 were considered to be a match. A color--color diagram was then plotted, as shown in Figure \ref{fig:clust_sep}, with (F356W$-$F444W) on the $y$-axis and (F090W$-$F200W) on the $x$-axis. Objects with $0.77<z_\mathrm{spec}<0.97$ primarily trace a sequence in this space bounded by $0.35<\mathrm{(F090W-F200W)}<1.64$ and $-0.40<\mathrm{(F356W-F444W)}<-0.26$. We adopt all objects inside this box as cluster members.
}

\begin{figure*}[ht!]
\plotone{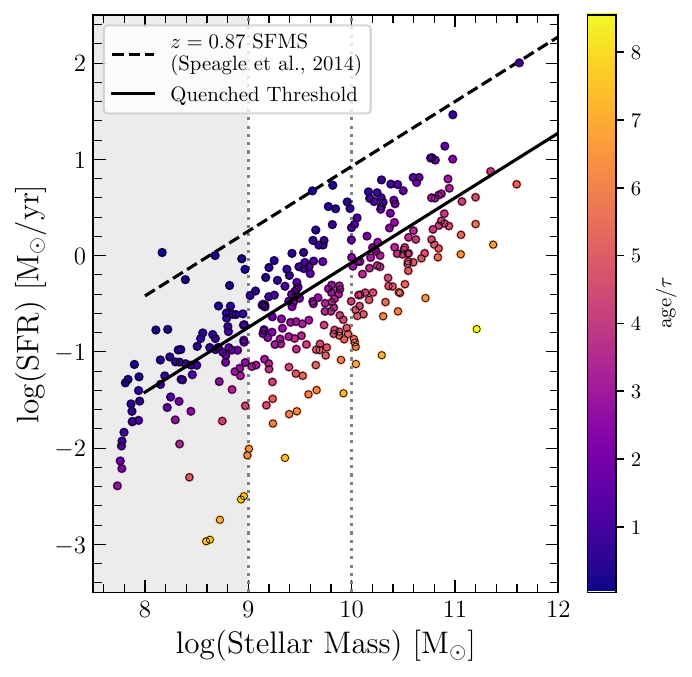}
\caption{The parameters $\log(\mathrm{SFR})$ vs. $\log(M_*)$ derived by SED fits of JWST/NIRCam photometry of El Gordo cluster members. The dashed black line shows the star-formation main sequence best fit from \citet{2014ApJS..214...15S} for $z=0.87$. The solid black line is 1~dex below the SFMS, representing the quenched threshold as described by \citet{2021MNRAS.500.4004D} and \citet{2023ApJ...954...98P}. The color represents the age/$\tau$ value for each galaxy according to the color bar on the right---as expected, galaxies with lower SFRs have older ages parameterized this way. The shaded region represents the stellar mass range in which we are incomplete, and the dotted grey lines represent the two mass bins we split the data into.
\label{fig:sfr_sm}}
\end{figure*}

We used \texttt{Bagpipes} \citep{2018MNRAS.480.4379C} to fit SEDs for all objects in both the parallel modules. The redshifts were fixed when performing this fitting to the $z_\mathrm{phot}$ measurements obtained with \texttt{EAZY}. A dust component with an $A_V$ range of $[0, 4]$ mag was included. An exponentially-declining star formation history (i.e., $\mathrm{SFR}(t)=\mathrm{SFR}_0\exp(-t/\tau)$) with an age range of $[0.01, 13.5]$ Gyr, a $\tau$ range of $[0.3, 10.0]$ Gyr, a metallicity range of $[0, 2.5]$ $Z_\odot$, and a \texttt{massformed} range of $[8, 12]$ $\mathrm{M}_\odot$ were used for this fitting. 
In order to find the best model for each galaxy, the fitting was performed several times with slightly different parameter ranges (i.e., priors), and the $\chi^2$ histograms for the various runs were compared. As another test, objects were fit with their $z_\mathrm{spec}$ value when this was available. This allowed us to optimize the other parameters while the redshifts were set to their true values before fitting the entire catalog using $z_\mathrm{phot}$ values. Varying the redshift to a range rather than fixing it was also tested, but comparison of the $z_\mathrm{phot}$ distributions to the $z_\mathrm{spec}$ distribution showed that the \texttt{EAZY} redshift estimates were more accurate than the ones made by \texttt{Bagpipes}. The outputs from this fitting which are most important for this study are the SFR and stellar mass ($M_\star$) values along with their $1 \sigma$ uncertainties. We took the SFR and $M_\star$ values to be the median posterior values from \texttt{Bagpipes} and the uncertainties to be the 16th--84th percentile ranges. Because photometric uncertainties make this selection technique less reliable for less massive galaxies, we only include galaxies with $\log(M_*/\mathrm{M}_\odot)>9$ in our analysis.

\begin{figure*}[bt!]
\plotone{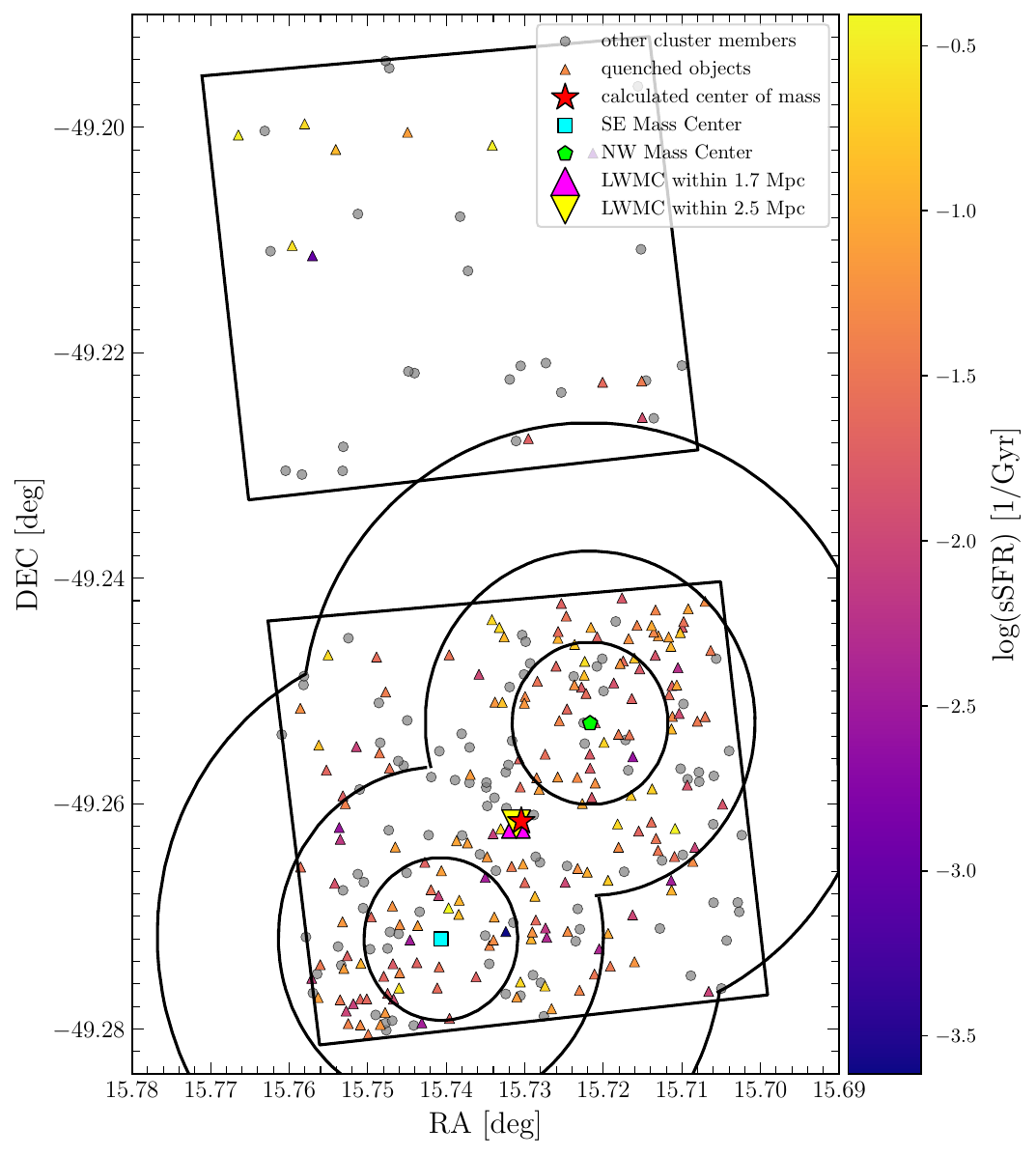}
\caption{Positions of cluster galaxies. Quenched objects are shown as  triangles with colors indicating specific SFR as shown on the color bar to the right of the plot. The grey circles mark the other (i.e., not quenched) cluster galaxies. The cyan square and green pentagon mark the southeast and northwest mass centers, respectively, and the pink and yellow triangles mark the luminosity-weighted mass centers (LWMCs) within 1.7 and 2.5\,Mpc \citep{2023ApJ...952...81F}. The red star marks the center of mass as computed using the Bagpipes stellar masses of the cluster objects in Module~B\null. The regions separated by black lines represent the three regions where quenched fraction was measured. The circles are centered on the mass centers and have radii 26\arcsec, 55\arcsec, and 96\arcsec.
\label{fig:annuli}}
\end{figure*}

\begin{figure}[bt!]
    \includegraphics[width=\linewidth]{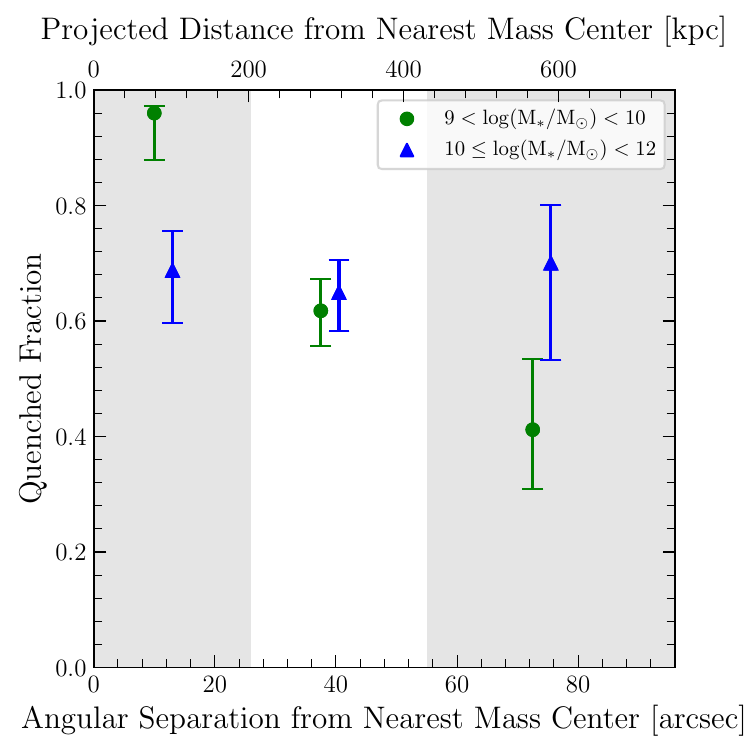}
    \caption{The quenched fraction of galaxies in El Gordo as a function of distance to the nearest mass center. Colored points show the quenched fractions within 3 angular separation bins from the nearest mass center. Angular separation from the nearest mass center is given in arcseconds on the bottom axis, and proper distance from the nearest mass center is given in kiloparsecs on the top axis. The sample is separated into $9 < \log{(M_*/\mathrm{M}_{\odot})} < 10$ (green circles), and $10 \leq \log{(M_*/\mathrm{M}_{\odot})} < 12$ (blue triangles). Shading indicates the boundaries between different regions.
    \label{fig:quenched_frac}}
\end{figure}

\section{Quenching in the El Gordo Cluster}
\subsection{Star Formation Rate vs.\ Stellar Mass}\label{secs1}
Star-forming galaxies show a strong correlation between SFR and $M_\star$ (e.g., \citealt{2004MNRAS.351.1151B}).  This relation is known as the ``star-formation main sequence'' (SFMS; \citealt{2007ApJ...660L..43N}). The SFMS evolves with redshift \citep[e.g.,][]{2007ApJ...660L..43N,2014ApJS..214...15S,Popesso2023} with galaxies with a given $M_\star$ having higher SFR at higher redshift. El Gordo members show a strong SFR--$M_\star$ correlation (Figure~\ref{fig:sfr_sm}), but nearly all cluster members have SFR smaller than the SFMS for field galaxies \citep{2014ApJS..214...15S} at $z=0.87$.  This is consistent with other observations that show galaxies in clusters have lower SFR than field galaxies of the same mass \citep{1980ApJ...236..351D,2012ApJ...746..188M,2022MNRAS.515.5479B}.

For this paper, we  adopt the definition of ``quenched" from both \citet{2019MNRAS.485.4817D} and \citet{2023ApJ...954...98P}. This definition states that quenched galaxies are those with SFRs more than 1~dex below the SFMS \citep{2014ApJS..214...15S, 2012ApJ...754L..29W}. \citet{2019MNRAS.485.4817D} did not distinguish between quenched and quiescent, and we will follow their lead, simply referring to objects with SFR at least 1~dex below the SFMS as ``quenched."
With this definition, 144 of the 229 galaxies in the cluster sample with $\log(M_*/\mathrm{M}_\odot)>9$ are quenched, giving a quenched fraction of 63\%.

\subsection{Quenched Galaxy Fraction vs.\ Distance}\label{secs2}

Insight into the quenching process may come from investigating the radial dependence of the quenched fraction. El Gordo has two main mass components located on either side of the overall center of mass. Figure~\ref{fig:annuli} shows the positions of all cluster members along with the various mass and luminosity centers, and Figure~\ref{fig:quenched_frac} shows the quenched fractions as a function of distance.

\begin{figure}[bt!]
    \includegraphics[width=\linewidth]{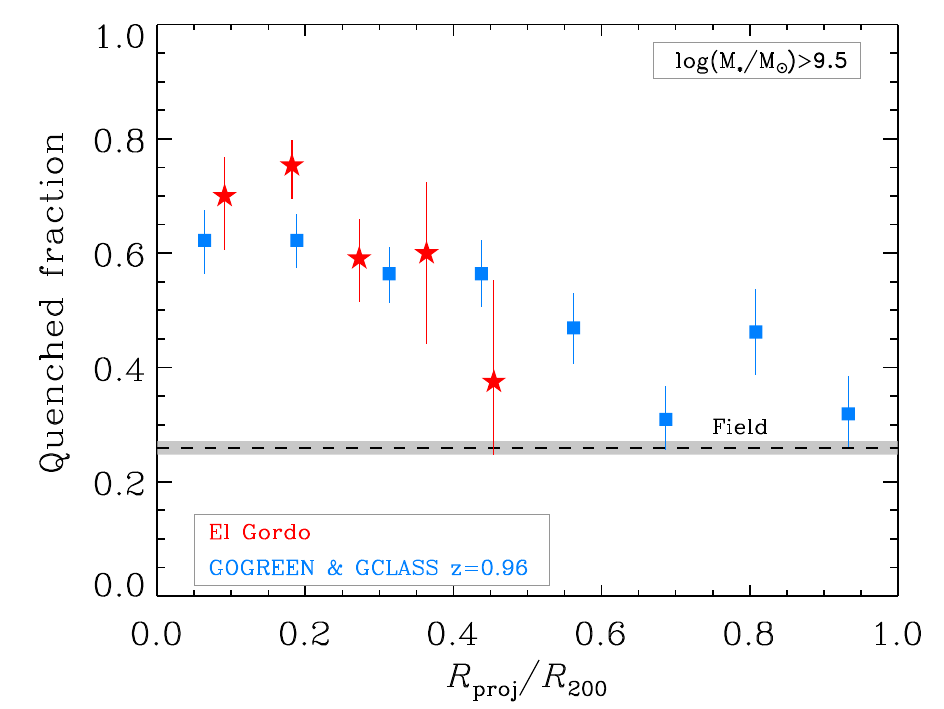}
    \caption{Quenched fraction compared to other clusters. The horizontal axis is the projected distance to the cluster mass center (nearest of the two mass centers, for El Gordo)} normalized to $R_{200}$, and all galaxies have $\log({M}_*/\mathrm{M}_{\odot})>9.5$. Red stars show measurements of the quenched fraction of galaxies in El Gordo, and blue squares show the quenched fractions for a sample of clusters from the GOGREEN and GCLASS surveys at $0.867<z<1.12$ ($\langle z\rangle=0.96$) \citep{Hewitt2025}.  The dashed black line shows the quenched fraction for field galaxies  \citet{2016ApJ...827L..25M} at $z=0.75$--1.0 and with ${M}_*>10^{9.4}$\,\Msol.
    \label{fig:QF_distance}
\end{figure}

The quenched fraction of low-mass ($9<\log(M_*/\mathrm{M}_\odot)<10$) galaxies in El Gordo decreases as angular separation from the nearest mass component increases, while high-mass ($10\leq\log(M_*/\mathrm{M}_\odot)<12$) galaxies show a near-constant quenched fraction (Fig.~\ref{fig:quenched_frac}). Specifically, the low-mass galaxies within 26\arcsec\ of a mass center are nearly all quenched,  while at $>$55\arcsec, fewer than half are quenched.
%
Figure~\ref{fig:QF_distance} compares El Gordo's quenched fraction as a function of normalized radius (El Gordo's virial radius is 1.75 Mpc \citep{2023A&A...672A...3D}, which corresponds to 220\arcsec) to that of less-massive  clusters at redshifts similar to El Gordo's \citep{Hewitt2025}. For a consistent comparison, all quenched fraction values in Figure \ref{fig:QF_distance} refer to galaxies with $\log(\mathrm{M}_*/\mathrm{M}_\odot)>9.5$. The quenched fraction in the comparison sample increases towards the cluster core, consistent with other studies from the literature \citep[e.g., ][]{2022MNRAS.515.5479B,vulcani13,haines15, Kawinwanichakij17,Pintos-Castro19}.
In El Gordo, the quenched fraction also increases with decreasing distances.


An alternate way to measure distance would be from the overall cluster mass center. This gives quenched fractions depending strongly on mass but little on distance with the lowest-mass galaxies having the highest quenched fractions across all distances. This is consistent with the findings of \citet{2016MNRAS.463.1916F, 2022MNRAS.515.5479B, 2013MNRAS.432..336W, 2012ApJ...761..142M}. However, these results are probably an artifact of the densest part of the clusters being 39\arcsec--46\arcsec\ from the overall mass center.

\subsection{Quenched Galaxy Fraction vs.\ Redshift}


\begin{figure}[ht!]
    \includegraphics[width=\linewidth]{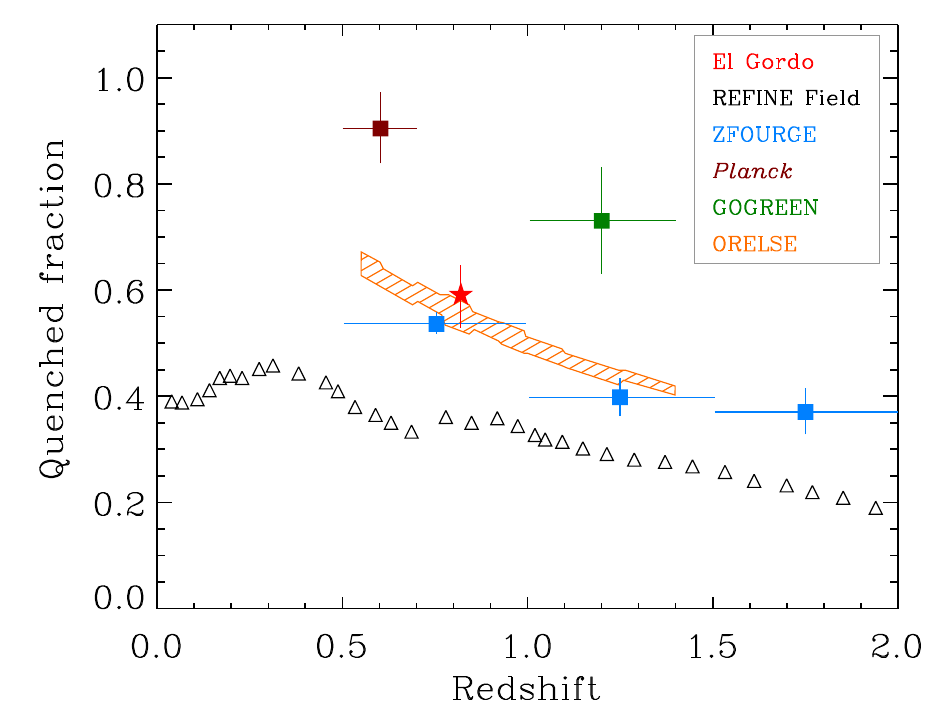}
    \caption{Quenched fraction for massive galaxies in El Gordo compared to galaxies in other clusters.  All comparisons are for galaxies with $10.25<\log(M_*/\mathrm{M}_\odot)<11$ and $R/R_{200}<0.5$. El Gordo is shown as a red star. The other symbols refer to galaxies from the field derived in the REFINE survey \citep[black open triangles; ][]{sarron21}, from the ZFOURGE group sample \citep[full blue squares; ][]{Papovich18,Straatman16}, from the ORELSE cluster sample \citep[orange hatched region; ][]{Lubin09,Straatman16}, from the {\it Planck} cluster sample at $0.5<z<0.7$ \citep[full maroon square;][]{vanderburg18}, and from the the GOGREEN cluster sample \citep[full green square;][]{vanderburg20,balogh17}.}
    \label{fig:QF_redshift}
\end{figure}

Figure~\ref{fig:QF_redshift} compares the quenched fractions of massive galaxies in El Gordo  with those measured in other clusters \citep{sarron21}  at $0.3<z<1.8$.
El Gordo's quenched 
fraction is $\sim$59\% within 110\arcsec\ ($=R_{200}/2$) of the nearest mass component. In general, groups and clusters have higher quenched fractions than field galaxies, and the difference increases at decreasing redshifts (e.g., \citealt{2023MNRAS.522.2297T,vanderburg20}). \citet{2024MNRAS.528.6329A} find that this higher quenched fraction in clusters extends to high-mass ($10<\log(M_*/\mathrm{M}_\odot)<11$) galaxies but that the dominant quenching mechanisms in these galaxies are the same as those for high-mass field galaxies, which they attribute to high-mass cluster galaxies having higher peak maximum circular dark matter halo velocities than those in the field. Our Figure \ref{fig:QF_redshift} is consistent with these findings in that it shows higher overall quenched fractions among $10.25<\log(M_*/\mathrm{M}_\odot)<11$ galaxies in clusters and groups than in field galaxies. While the quenched fraction in El Gordo is higher than the field fraction, it is lower than that measured in other $z\sim1$ clusters

\section{Discussion}
The results in Figure~\ref{fig:quenched_frac} show that near the El Gordo mass centers, galaxies with $M_*\lesssim 10^{10}$~\Msol\ are quenched at a higher rate than more massive galaxies.
One plausible quenching mechanism is ram-pressure stripping, first proposed by \citet{1972ApJ...176....1G} to explain the lack of gas-rich galaxies within clusters. The idea is that galaxies experience ICM ``winds" during infall which can strip them of their gas halos if the winds are strong enough to overcome the gravitational attraction between the stellar and gas disks \citep{1972ApJ...176....1G,2006ApJ...647..910H}. Many studies (e.g., \citealt{10.1111/j.1365-2966.2012.20983.x,2013MNRAS.432..336W,2015ApJ...808L..27W,2014MNRAS.442.1396W,2015MNRAS.454.2039F}) have confirmed that stripping is a highly mass-dependent process that is more effective at lower stellar masses. Due to the fact this process is heavily reliant on hot gas, the quenched fraction would be expected to be higher in areas of the cluster with more hot gas. However, this does not appear to be the case for El Gordo, which has a strong peak in x-ray emission near the SE mass peak \citep{2012ApJ...748....7M}. From Figure \ref{fig:annuli}, it can be seen that the quenched fraction is distributed relatively evenly around the SE and NW mass centers. Among the whole $9\leq \log(M_*)\leq12$ population of El Gordo members, we find quenched fractions of $65.3\%^{+5.10}_{-5.93}$ and $72.4\%^{+4.48}_{-5.63}$ within 42.6\arcsec\ of the SE and NW mass centers, respectively. This 1$\sigma$ difference is not statistically significant and not consistent with ram-pressure stripping.


Another possible quenching mechanism is strangulation, especially in the middle radii where the ICM density is lower, and ram pressure might not be as effective. \citet{2000ApJ...540..113B} presented a model to explain the gradual cluster-centric gradient of quenching in galaxy clusters in which the SFRs of accreted galaxies gradually decline over many Gyr, proposing that the cluster-centric gradient results from the strong correlation between radius and accretion times. This model also explains the decreased SFR of galaxies in the cluster outskirts, as far out as $R/R_\mathrm{vir}\sim2$, compared to those in the field as being due to cluster members being disturbed during major merger events and sent into highly-eccentric loose orbits. \citet{2013MNRAS.432..336W} built off this idea, suggesting a ``delayed-then-rapid" quenching scenario in which there is a 2--4~Gyr delay in quenching after infall followed by a rapid decline in SFR over $<0.8$~Gyr. Both of these timescales are shorter for more massive satellite galaxies, so this process could help explain the more gradual radial decline seen in the quenched fraction of the low-mass El Gordo galaxies in this study.

The flat quenched fraction trend for high-mass galaxies is consistent with internal quenching mechanisms, such as AGN feedback, which are independent of cluster-centric distance (e.g., \citealt{2013MNRAS.428.3306W}). The transition in the quenched fraction's dependence on radius near $M_*\sim10^{10}~\mathrm{M}_\odot$ found in this work is consistent with other studies that show a transition in quenching mechanisms at this mass (e.g., \citealt{2014MNRAS.439.3564C,2015ApJ...810...90L,2017ApJ...841L..22G}).

\section{Conclusions}
We have identified a sample of 283 galaxies in the El Gordo cluster to investigate the impact of the cluster environment on quenching.
For the selected galaxies, we identified the quenched subset as those with a $\geq$1~dex offset from the SFMS and measured the quenched fraction as a function of angular separation from the nearer of El Gordo's two mass centers. The quenched fraction for $9<\log({M}_*/\mathrm{M}_\odot)<10$ galaxies displays a dependence on the angular separation from the nearest mass component, while the $10\leq\log({M}_*/\mathrm{M}_\odot)<12$ galaxies do not.

The sample completeness in this study reaches $\log({M}_*/\mathrm{M}_\odot)\sim 9$. One way to gain a better understanding of quenching in El Gordo is to extend this completeness to masses ${<}10^9\,\mathrm{M}_\odot$. Deeper NIRCam imaging would accomplish this, particularly in  short wavelength-bands where the effects of quenching are expected to be more apparent, and quenched galaxies are fainter.

\facilities{HST and JWST Mikulski Archive \url{https://archive.stsci.edu}}


\software{Astropy \citep{2013A&A...558A..33A,2018AJ....156..123A,2022ApJ...935..167A}; SourceExtractor \citep{1996A&AS..117..393B}; EAZY \citep{2008ApJ...686.1503B}; Bagpipes \citep{2018MNRAS.480.4379C}}
\section{Acknowledgments}
\begin{acknowledgements}
    This work is based on observations made with the NASA/ESA/CSA James Webb Space
    Telescope. The data were obtained from the Mikulski Archive for Space Telescopes
    (MAST) at the Space Telescope Science Institute, which is operated by the
    Association of Universities for Research in Astronomy, Inc., under NASA contract
    NAS 5-03127 for JWST. These observations are associated with JWST program 1176.
    RAW, SHC, and RAJ acknowledge support from NASA JWST Interdisciplinary
    Scientist grants NAG5-12460, NNX14AN10G and 80NSSC18K0200 from GSFC. TMC is grateful for support from the Beus Center for Cosmic Foundations.
\end{acknowledgements}




\bibliography{quenched_fraction}{}
\bibliographystyle{aasjournal}



\end{document}